
\magnification=\magstep1
\overfullrule=0pt
\setbox0=\hbox{{\cal W}}

\def\w{{\cal W}}

\setbox9=\hbox{{\cal S}}

\def\sw{{\cal SW}}
\def\n{{\cal N}}

\def\B{{\cal B}}

\def\F{{\cal F}}

\def\ob{\bigl (}
\def\cb{\bigr )}
\def\lb{\lbrack}
\def\rb{\rbrack}
\def\de{\partial}

\def\q#1{\lb#1\rb}
\def\mn{\medskip\smallskip\noindent}
\def\sn{\smallskip\noindent}
\def\bn{\bigskip\noindent}

\font\covar=cmssi10 scaled \magstep0

\def\coc{{\hbox{\covar C}}}

\def\cowww{C^{\Phi}_{\Phi\Phi}}

\font\extra=cmss10 scaled \magstep0 \font\extras=cmss10 scaled 750

\setbox1 = \hbox{{{\extra R}}}
\setbox2 = \hbox{{{\extra I}}}
\setbox3 = \hbox{{{\extra C}}}

\setbox4=\hbox{{{\extra Z}}}
\setbox5=\hbox{{{\extras Z}}}
\setbox6=\hbox{{{\extras z}}}
\setbox7=\hbox{{{\extra N}}}
\def\Z{{{\extra Z}}\hskip-\wd4\hskip 2.5 true pt{{\extra Z}}}

\def\zs{{{\extras z}}\hskip-\wd6\hskip 1.7 true pt{{\extras z}}}
\def\EN{{{\extra N}}\hskip-\wd7\hskip-1.5 true pt{{\extra I}}\hskip-\wd2
\hskip1.5 true pt\hskip\wd7}
\def\Zed{\hbox{{\extra\Z}}}

\def\zeds{\hbox{{\extras\zs}}}
\def\N{\hbox{{\extra\EN}}}

\def\nota{2}
\def\rep{3}
\def\res{4}
\def\summa{5}
\def\elfh{{11\over2}}
\def\nh{{9\over2}}
\def\sh{{7\over2}}
\def\fh{{5\over2}}
\def\dh{{3\over2}}
\def\eh{{1\over2}}

\def\bbss{1}
\def\bai{2}
\def\bal{3}
\def\bpz{4}
\def\bershadsky{5}
\def\blg{6}
\def\blumenhagen{7}
\def\supwir{8}
\def\blm{9}
\def\bou{10}
\def\cap{11}
\def\wirrep{12}
\def\eichenherr{13}
\def\fal{14}
\def\fateev{15}
\def\fig{16}
\def\jfss{17}
\def\mfl{18}
\def\goddard{19}
\def\gko{20}
\def\hornfeck{21}
\def\ich{22}
\def\inami{23}
\def\ka1{24}
\def\kau{25}
\def\komata{26}
\def\mussardoi{27}
\def\mussardoii{28}
\def\nahm{29}
\def\scheunert{30}
\def\rva{31}
\def\zam{32}
\font\HUGE=cmbx12 scaled \magstep4
\font\Huge=cmbx10 scaled \magstep4
\font\Large=cmr12 scaled \magstep3
\font\MLarge=cmti12 scaled \magstep3
\font\large=cmr17 scaled \magstep0
%
%
\nopagenumbers
\pageno = 0
\centerline{\HUGE Universit\"at Bonn}
\vskip 10pt
\centerline{\Huge Physikalisches Institut}
\vskip 2cm
\centerline{\Large Representations of {\MLarge N}$\,$=$\,$1 Extended}
\vskip 6pt
\centerline{\Large Superconformal Algebras}
\vskip 6pt
\centerline{\Large with Two Generators}
\vskip 0.8cm
\centerline{\large W.\ Eholzer, A.\ Honecker, R.\ H\"ubel}
\vskip 1.5cm
\centerline{\bf Abstract}
\vskip 15pt
\noindent
In this paper we consider the representation theory of
$N=1$ Super-$\w$-algebras with two generators for conformal
dimension of the additional superprimary field between two and
six. In the superminimal case our results coincide with the
expectation from the ADE-classification. For the parabolic
algebras we find a finite number of highest weight
representations and an effective central charge $\tilde c = \dh$.
Furthermore we show that most of the exceptional algebras lead
to new rational models with $\tilde c > {3\over2}$. The
remaining exceptional cases show a new `mixed' structure.
Besides a continuous branch of representations discrete values
of the highest weight exist, too.
\vfill
\settabs \+&  \hskip 110mm & \phantom{XXXXXXXXXXX} & \cr
\+ & Post address:                       & BONN-HE-92-28   & \cr
\+ & Nu{\ss}allee 12                     & hep-th/9209030  & \cr
\+ & W-5300 Bonn 1                       & Bonn University & \cr
\+ & Germany                             & September 1992  & \cr
\+ & e-mail: ralf@mpim-bonn.mpg.de       & ISSN-0172-8733  & \cr
\eject
\pageno=1
\footline{\hss\tenrm\folio\hss}
%
%
\leftline{\bf 1. Introduction}
\mn
One of the most interesting questions in two-dimensional conformal quantum
field theory is the classification of all rational models. So far no
general classification schemes leading to a complete list of all RCFTs
have been developped. One hopes to describe all
rational models as minimal models of the symmetry algebra of a
conformally invariant theory, which therefore contains the Virasoro
algebra as a subalgebra. Consequently a reasonable ansatz is
the construction of extensions of the conformal algebra,
$\w$-algebras, and the study of their highest weight representations.
\sn
After the initial work of Zamolodchikov in 1985 $\q{\zam}$ several methods
have been found to construct $\w$-algebras. Besides the GKO-construction
$\q{\gko}\q{\bal}\q{\blg}\q{\bbss}\q{\bai}$ and the free field approach
$\q{\fateev}\q{\fal}$ a method constructing the algebra
explicitly exists $\q{\blm}\q{\kau}$, which we call the Lie bracket approach.
Another equivalent method is the conformal bootstrap
$\q{\bpz}\q{\fig}\q{\bou}$. Starting with the results of
$\q{\blm}\q{\rva}$ we investigated the representation theory of
$\w(2,d)$-algebras $\q{\wirrep}$.
\sn
With the methods cited above several $N=1$
supersymmetric $\w$-algebras have been constructed
$\q{\jfss}\q{\inami}\q{\blumenhagen}$.
Recently we constructed a lot of new of $N=1$ $\sw$-algebras with two
and three generators using the non-covariant Lie bracket approach
$\q{\supwir}$. These results led to a better understanding of the
classification of $N=1$ $\sw$-algebras. In this paper we use the
same methods as in $\q{\wirrep}$ to investigate the representation
theory of the $\sw(\dh,\delta)$-algebras with conformal dimension
of the additional generator between two and six. We stress the
fact that our methods yield necessary conditions for the existence
of consistent highest weight representations.
\sn
In the next chapter we recall the notions and the results of
the construction of $N=1$ $\sw$-algebras. We proceed with a
chapter concerning the representation theory of $\sw$-algebras
and discuss some new problems arising in the Ramond sector.
In chapter four we present the explicit results of our calculations.
We end with a summary in the last chapter.
\bn
\leftline{\bf \nota. Notations and previous results}
\mn
Let $\F$ be the algebra of local chiral fields of conformal field theory
defined on a two-dimensional compactified spacetime. The
requirement of invariance under rational conformal transformations induces
a grading of $\F$ by the conformal dimension. We assume that $\F$ is
spanned by the quasiprimary fields and derivatives thereof. Furthermore
$\F$ carries the (non-associative) operation of forming normal ordered
products.
\noindent Let $\phi(z) := \sum_{{\scriptscriptstyle{n-d(\phi)\in}}\zeds}
z^{n-d(\phi)} \phi_n $ be the Fourier
decomposition of the chiral field $\phi(z)$ with conformal dimension $d(\phi)$
 and
call $\phi_n$ the modes of $\phi$. It is well known that the modes $\lbrace
L_n | n\in\Zed \rbrace$ of the energy momentum tensor satisfy the Virasoro
algebra. A primary field $\phi$ obeys the following commutation relations:
$$\lb L_n,\phi_m\rb = \ob m-(d(\phi)-1)n\cb \phi_{n+m} \eqno({\rm \nota.1})$$
If (\nota.1) is valid only for $n\in\lbrace -1,0,1\rbrace$, $\phi$ is a
quasiprimary field.
\sn
W.\ Nahm showed that the requirements of locality and invariance under rational
conformal transformations determine the commutator of two chiral quasiprimary
 fields
up to some structure constants which are determined by additional
dynamical principles (i.e.\ Jacobi identities) $\q{\nahm}$. Let
$\lbrace\phi_i\!\mid\! i\!\in\! I\rbrace$ be a set of quasiprimary chiral
fields
with conformal dimensions $d(\phi_i) \in \N /2$ which together with their
derivatives span $\F$. The commutator of the modes of the quasiprimary fields
is given by
\sn
$$\lb\phi_{i,m},\phi_{j,n}\rb_{\pm}=\sum_{k\in I}C^k_{ij}\,p_{ijk}(m,n)
\phi_{k,m+n} + d_{ij}\,\delta_{n,-m}{n+d(\phi_i)-1\choose
2d(\phi_i)-1}\eqno{{\rm(\nota.2)}}$$
$d_{ij}$ describes the normalization of the two point functions and
$C^k_{ij}$ are coupling constants. $p_{ijk}$ are universal polynomials
depending only on the conformal dimensions of the fields involved (for
details see $\q{\blm}$).
\sn
The Fourier modes of the normal ordered product of two chiral fields are
defined by
$$N(\phi,\psi)_n := (-1)^{4 d(\phi) d(\psi)}
 \sum_{k<d(\psi)}\!\phi_{n-k}\psi_k\,+
\sum_{k\geq d(\psi)}\!\psi_k\phi_{n-k}   \eqno({\rm \nota.3})$$
To make use of the commutator formula (\nota.2) one has to consider the
quasiprimary projection of (\nota.3):
$$\n(\phi_j,\partial^{n}\phi_i) := N(\phi_j,\de^{n}\phi_i)
-\bigl( \sum_{r=1}^{n}\alpha^r_{ij} \de^r N(\phi_j,\de^{n-r}\phi_i)
 + \!\!\sum_{k} \beta^k_{ij}(n)
C^k_{ij} \partial^{\gamma^k_{ij}(n)}\phi_k\bigr)
  \eqno( {\rm \nota.4})$$
$\alpha^r_{ij},\beta^k_{ij}(n),\gamma^k_{ij}(n)$ are polynomials depending
only on the conformal dimensions of the fields involved $\q{\blm}$.
This definition yields a quasiprimary field of conformal
dimension $ d(\phi_i)+d(\phi_j)+n $.
\sn
As the number of quasiprimary fields in $\F$ is infinite we introduce
the notion of `simple' fields. Let
${\cal B} = \lbrace \phi_i \mid i \in I \rbrace$
be a subset of the quasiprimary fields of $\F$. $\B$ generates $\F$
if it is possible to obtain a (vectorspace) basis of $\F$ out of
$\B$ using the operations of forming derivatives and normal ordered products.
If the fields in $\B$ are orthogonal to all normal ordered products in $\F$
we call them `simple' fields.
\mn
The super Virasoro algebra is the extension of the Virasoro algebra by a
primary simple field $G(z)$ of conformal dimension $\dh$.
It is a Lie superalgebra
and exists for generic values of the central charge $c$.
With the standard normalization $d_{\phi\phi} = {c\over{d(\phi)}}$
of a quasiprimary field $\phi$ with conformal dimension $d(\phi)$
the commutation relations take the following form:
$$\eqalign{
\lb L_m,L_n\rb &= (n-m)L_{m+n}+{\textstyle{c\over12}}(n^3-n)\delta_{n+m,0} \cr
\lb L_m,G_n\rb &= (n-{\textstyle{1\over2}}m)G_{m+n} \cr
{\lb G_m,G_n\rb}_{+} &= 2L_{m+n} +
{\textstyle{c\over3}}(m^2-{\textstyle{1\over4}})\delta_{m+n,0}
} \eqno( {\rm \nota.5}) $$
This algebra is the central extension of the algebra
formed by the generators of superconformal transformations in
superspace $Z$, consisting of the points $(z,\theta)$ where
$\theta$ is a Grassmannian variable.
A field $\Phi(Z) = \phi(z) + \theta \psi(z)$ is called super primary with
superconformal dimension $\delta=d(\Phi)$, iff $\phi$ and $\psi$ are Virasoro
 primaries
with conformal dimension $\delta$ resp.\ $\delta+\eh$ and satisfy the following
commutation relations
$$\eqalign{
{\lb G_n,\phi_m\rb}_{\pm} &= \coc^{\psi}_{G\phi}
p_{\dh,\delta,\delta+\eh}(n,m)\ \psi_{n+m}
= \coc^{\psi}_{G\phi} \psi_{n+m} \cr
{\lb G_n,\psi_m\rb}_{\pm} &= \coc^{\phi}_{G\psi}
p_{\dh,\delta+\eh,\delta}(n,m)\ \phi_{n+m}
= {\coc^{\phi}_{G\psi}\over{2\delta}}(m-(2\delta-1)n) \phi_{n+m}
} \eqno( {\rm \nota.6} ) $$
$\coc^{\psi}_{G\phi}$ and $\coc^{\phi}_{G\psi}$ are special known
structure constants $\q{\supwir}$.
If $\phi$ and $\psi$ are quasiprimaries and (\nota.6)
is fulfilled only for $n\in\lbrace -\eh,\eh \rbrace$, $\Phi$ is called a
super quasiprimary.
\sn
{\bf Definition:} Let $\B = \lbrace G,L,\phi_1,\psi_1,\ldots,\phi_n,\psi_n
\rbrace $ be a set of simple fields with the additional property
that $\Phi_i = \phi_i + \theta \psi_i, i=1,\ldots,n$ are super primaries.
The associative field algebra
generated by $\B$ is called a $\sw\ob\dh,d(\Phi_1),\ldots,d(\Phi_n)\cb$.
\sn
For a $\sw(\dh,\delta_1,\ldots,\delta_n)$-algebra
to be consistent it is necessary
that all Jacobi identities involving the additional primary fields are
fulfilled. In $\q{\supwir}$ we described
in detail how to construct the algebras using the Lie bracket
approach. Let us recall the results of the construction of
$\sw(\dh,\delta)$-algebras.
The central charge and the conformal dimensions
of the super primaries of the minimal models of
the super Virasoro algebra are given
by the following relations
$\q{\bershadsky}\q{\eichenherr}\q{\goddard}\q{\mussardoi}\q{\mussardoii}$:
$$\eqalign{
{\textstyle c(p,q)} &= {\textstyle{3\over2}}\ob
{\textstyle{1\! -\! {{2(p-q)^2}\over{pq}} }}\cb
\ p,\! q\!\in\N,{\rm gcd}(p,\! q)\!=\!1,p\! +\! q\! \in\! 2\,\N
\vee p,\! q\!\in\! 2\,\N,
{\rm gcd}({\textstyle{p\over2},\! {q\over2}})\!=\!1,
{\textstyle {p\over2}\!+\!{q\over2}}\!\not\in\! 2\,\N \cr
{\textstyle h(r,s)} &= {\textstyle{ {{(rp-qs)}^2-{(p-q)}^2}\over{8pq} } }
+{\textstyle {1-{(-1)}^{r+s}\over32}}
 \ \ \ 1 \le r \le q-1 \ , \  1 \le s \le p-1
 }  \eqno({\rm \nota.7}) $$
$r+s$ even yields representations in the Neveu-Schwarz sector and $r+s$ odd
representations in the Ramond sector.
In perfect analogy to $\w(2,d)$-algebras
the $c$-values for which the $\sw(\dh,\delta)$-algebras are consistent
can be divided in five classes.
\sn
\item{$\bullet$} generically existing $\sw(\dh,\delta)$-algebras
\item{$\bullet$} algebras existing for superminimal values of the central
charge and therefore being related to the models of the ADE-classification of
A.\  Cappelliet al. $\q{\cap}$
\item{$\bullet$} algebras existing for central charges
of the form $c=c(1,s)$,\ $2\leq s\in\N$
\item{$\bullet$} parabolic algebras related to
degenerate models of the super Virasoro algebra
\item{$\bullet$} exceptional algebras which cannot be put in any other class
\mn
In the following table we recall the concrete values of the central charge
for the $\sw$-algebras $\sw(\dh,\delta)$ with $\dh\leq\delta\leq 6$ and
their classification into the different classes.
\bn
\centerline{\vbox{ \offinterlineskip
\def\tablespace{ height2pt&\omit&&\omit&&\omit&&\omit&&\omit&&
 \omit&&\omit&&\omit&&\omit&&\omit&\cr }
\def\tablespaceo{ height2pt&\omit&&\omit&&\omit&&\omit&&\omit&& \multispan3
 &&\omit&&\omit&&\omit&\cr }
\def\tablerule{ \tablespace\noalign{\hrule}\tablespace}
\def\tablerulesm{ \noalign{\hrule}\tablespace}
\hrule\halign{&\vrule#&\strut\hskip5pt\hfil#\hfil\hskip5pt\cr
\tablespace
& $\delta$ && \hskip-5pt && $c$ && series && \hskip-8pt &&  $c$ && series &&
 \hskip-8pt &&  $c$ && series &\cr
\tablespace\noalign{\hrule}\tablespace
\tablespace\noalign{\hrule}\tablespaceo
\tablespaceo
& $2$ && \hskip-5pt && $-{6\over5}$ ($\cowww=0$) && $(A_3,D_6)$ && \hskip-8pt
&&\multispan3 generic ($\cowww \ne 0$)  &&  \hskip-8pt && \omit && \omit  &\cr
\tablespaceo\tablespaceo\noalign{\hrule}\tablespace\tablerulesm\tablespace
& ${5\over2}$ && \hskip-5pt && $-{5\over2}$ && $(1,s)$ && \hskip-8pt &&
 ${10\over7}$ && $(D_8,E_6)$ &&  \hskip-8pt &&\omit && \omit &\cr
\tablespace\tablerule\tablerulesm\tablespace
& $3$ && \hskip-5pt && $-{27\over7}$  && $(A_3,D_8)$ && \hskip-8pt &&
 ${5\over4}$ && $(A_7,D_4)$ &&  \hskip-8pt && $-{45\over2}$  && parabolic &\cr
\tablespace\tablerule\tablerulesm\tablespace
& ${7\over2}$ && \hskip-5pt && ${7\over5}$ && $(A_9,E_6)$ && \hskip-8pt &&
 $-{17\over11}$ && exceptional &&  \hskip-8pt && \omit && \omit &\cr
\tablespace\tablerule\tablerulesm\tablespace
& $4$ && \hskip-5pt && $-{20\over3}$ && $(A_3,D_{10})$ && \hskip-8pt &&
 $-{21\over2}$ && parabolic &&  \hskip-8pt && \omit && \omit &\cr
\tablespace\tablespace
& \omit && \hskip-5pt && $-{185\over4}$  &&  exceptional && \hskip-8pt && $-13$
 &&  exceptional &&  \hskip-8pt && $-{120\over13}$ &&  exceptional &\cr
\tablespace\tablerule\tablerulesm\tablespace
& ${9\over2}$ && \hskip-5pt && $-{81\over10}$ && $(1,s)$ && \hskip-8pt &&
 ${4\over11}$ && $(D_{12},E_6)$ &&  \hskip-8pt && $-{69\over2}$  && parabolic
 &\cr
\tablespace\tablerule\tablerulesm\tablespace
& $5$ && \hskip-5pt && $-{105\over11}$ && $(A_3,D_{12})$ && \hskip-8pt && \omit
 && \omit && \hskip-8pt && \omit && \omit &\cr
\tablespace\tablerule\tablerulesm\tablespace
& ${11\over2}$ && \hskip-5pt && $-{5\over13}$ && $(D_{14},E_6)$ && \hskip-8pt
&& ${10\over7}$ && $(A_{13},E_6)$ && \hskip-8pt && ${11\over40}$ &&
 $(A_{15},E_8)$&\cr
\tablespace\tablespace
& \omit &&  \hskip-5pt && $-{705\over8}$ && exceptional && \hskip-8pt &&
 $-{155\over19}$ && exceptional && \hskip-8pt && \omit && \omit &\cr
\tablespace\tablerule\tablerulesm\tablespace
& $6$ && \hskip-5pt && $-{162\over13}$  && $(A_3,D_{14})$ && \hskip-8pt &&
 ${27\over20}$ && $(A_7,D_6)$ && \hskip-8pt && $-{93\over2}$  && parabolic &\cr
\tablespace\tablespace
& \omit && \hskip-5pt && $-{33\over2}$ && parabolic && \hskip-8pt && $-18$ &&
 exceptional && \hskip-8pt && $-{2241\over20}$ &&  exceptional &\cr
\tablespace\tablespace}\hrule}}
\bn
\leftline{\bf \rep. Representations of $\sw(\dh,\delta)$ algebras}
\mn
Per definition $\sw$-algebras contain fermionic fields, i.e.\ fields
with half-integer conformal dimension. In the representation theory of these
algebras we have to distinguish between the Neveu-Schwarz and the Ramond
sector. From the mathematical point of view fermionic fields are defined on
a double sheet covering space of the complex plane and therefore have
nontrivial monodromy properties. In the Neveu-Schwarz (NS) and
Ramond (R) sector a fermionic field
$\Omega(z)$ obeys the following transformation laws under the transformation
$z\rightarrow e^{2\pi i} z$:
$$\eqalign{\Omega(e^{2\pi i} z) &= \Omega(z)\ \ \ \ \ {\rm NS\ sector}\cr
\Omega(e^{2\pi i} z) &= -\Omega(z)\ \ \ {\rm R\ sector}}$$
Consequently fermionic fields have integer modes in the Ramond sector and
half-integer modes in the Neveu-Schwarz sector.
\mn
In the following we will consider $\sw$-algebras with one additional
generator $\Phi=\phi+\theta\psi$. Let w.l.o.g.\ $\delta$ be an integer
and denote the algebra $\sw(\dh,\delta)$ by $\cal A$.
\sn
{\bf Definition:} The abelian two dimensional subalgebra $\cal C$ of
$\cal A$ generated by the zero modes of the bosonic
simple fields ($L_0,\phi_0$) is called the Cartan subalgebra of $\cal A$.
The subalgebra of $\cal A$ which is generated by the zero modes of all fields
is called the `horizontal' subalgebra.
\sn
{\bf Definition:} A linear representation ($V,\lambda$), $\lambda\in{\cal
 C}^{*}$
of $\cal A$ is a highest weight representation (HWR) iff the complex
vectorspace $V$ contains a nonzero unique cyclic
vector $\mid h,w\,\rangle$ with the following properties:
$$\eqalign{
L_0\ \mid h,w\,\rangle &= \lambda(L_0) \mid h,w\,\rangle\ =\ h \mid
 h,w\,\rangle\cr
\phi_0\ \mid h,w\,\rangle &= \lambda(\phi_0) \mid h,w\,\rangle\ =\ w \mid
 h,w\,\rangle\cr
L_n\ \mid h,w\,\rangle &= G_n\ \mid h,w\,\rangle = \phi_n\ \mid h,w\,\rangle
= \psi_n\ \mid h,w\,\rangle = 0\ \ \ \forall n < 0\cr
}\eqno({\rm \rep.1})$$
The representation module $V$ is spanned by the vectors which are obtained
by applying the positive modes of the simple fields to the highest weight
vector (HWV)  $\mid h,w\,\rangle$.
\mn
In the Neveu-Schwarz sector the representation theory of $\sw$-algebras
is very similar to that of ordinary fermionic $\w$-algebras.
But in the Ramond sector it is more complicated since here the zero modes
of the fermionic simple fields are involved.
First of all we establish that the ground state in the representation module
generally possesses a four-fold degeneracy because the vectors
$\mid h,w\,\rangle,\, G_0 \!\mid h,w\,\rangle,
\, \psi_0 \!\mid h,w\,\rangle,\, \psi_0 G_0 \!\mid h,w\,\rangle$
have all the same $L_0$-eigenvalue $h$. It is easy to see that there are no
more linearly independent vectors with $L_0$-eigenvalue $h$. In several cases
we checked explicitly that e.g.\ the vector $\psi_0 G_0\!\mid h,w\,\rangle$
is not proportional to the HWV $\mid h,w\,\rangle$.
The investigation of the representation theory of the horizontal subalgebra
should lead to a better understanding of the phenomenon of the ground state
degeneracy. For practical calculations this degeneracy has the consequence
that in the evaluation of Jacobi identities (see below) the correlators
$\langle\,h,w \mid G_0 \mid h,w\,\rangle,
\, \langle\,h,w \mid \psi_0 \mid h,w\,\rangle,
\, \langle\,h,w \mid \psi_0 G_0 \mid h,w\,\rangle$
appear as additional unknowns. Therefore we need in general one more linearly
independent condition than in the Neveu-Schwarz sector to restrict the
highest weight to a finite number of values (we evaluated only Jacobi
identities which contained an even number of fermionic fields
so that only the last one of the above correlators appeared).
\mn
The second difficulty arising in the Ramond sector is the definition of the
quasiprimary normal ordered product  $\n(\phi,\psi)$.
The standard definition (\nota.4) fails in the Ramond sector if
$\psi$ is a fermionic field $\q{\ich}$.
In the practical calculations we were able to
surround this problem by using
the commutator formula of W.\ Nahm (\nota.2). At first one avoids the
occurence of NOPs of the form
$\n(\phi,\psi)$ with $d(\phi)\in\N$ and $d(\psi)\in\N+\eh$
by choosing a special basis in the space of quasiprimary fields. It is possible
to choose a basis where only NOPs of fermionic with bosonic fields occur with
the bosonic field being the second entry of $\n(.,.)$.
However, in this basis NOPs of two fermionic fields still appear.
As an example consider the basis of quasiprimary fields
of conformal dimension $4$ constructed out of $L$ and $G$ which is given by
$\n(L,L),\n(G,\de G)$.
To handle the second NOP we use the commutator formula (\nota.2) to express
it by NOPs which can be evaluated with formula (\nota.4).
With (\nota.2) we obtain
$$\eqalign{
{\lb G_m\,,\,\n(G,L)_n \rb}_{+} =\ &p_{\dh,\sh,2}(m,n)\ C^L_{G\n(G,L)}\
L_{m+n}\ +\cr
&p_{\dh,\sh,4}(m,n)\ob C^{\n(L,L)}_{G\n(G,L)}\ \n(L,L)_{m+n}
+ C^{\n(G,\de G)}_{G\n(G,L)}\ \n(G,\de G)_{m+n}\cb
}\eqno({\rm\rep.3})$$
Inserting the polynomials and structure constants $\q{\ich}$ yields ($m = 0$):
$$ \n(G,\de G)_n = 2\ob G_0\n(G,L)_n + \n(G,L)_n G_0 - 2\,\n(L,L)_n
+ {\textstyle{1\over 48}}(4c+21)\,L_n\cb
\eqno({\rm\rep.4})$$
Analogously one obtains formulae for the other NOPs with two fermionic
fields. With this procedure we have been able to calculate the correct
 $h$-values
for the HWRs in the Ramond sector at the expense of computer time.
\mn
Let us finally describe in short the methods we used to obtain the relevant
HWRs.From the generically existing $\sw(\dh,\delta)$-algebras we considered
only the case $\delta=2$ because in all other cases the algebras are Lie
superalgebras for which the representation
theory is well understood $\q{\scheunert}$.
In complete analogy to the generically existing $\w(2,d)$-algebras
we determined in the continuum of representations for special $c$-values
rational models by constructing null fields. We stress that the evaluation
of Jacobi identities on the HWV did not yield any restrictions in this case.
\sn
The $\sw(\dh,\delta)$-algebras for $\delta\geq\fh$ exist only for discrete
values of the central charge. Here the evaluation of Jacobi identities
leads to restrictions of the highest weight. We considered the
following three- and four-point functions (w.l.o.g. $\delta \in\N$):
\sn
$$\eqalign{
0 &=\ \langle\,h\,,\,w\,\mid\lb\lb\phi_{-n},\phi_{-m}
\rb,\phi_{n+m}\rb_{cycl.}\mid\,h\,,\,w\,\rangle\cr
0 &=\ \langle\,h\,,\,w\,\mid\lb\lb\phi_{-n},\psi_{-m}
\rb_\pm,\psi_{n+m}\rb_{\pm,cycl.}\mid\,h\,,\,w\,\rangle\cr
0 &=\ \langle\,h\,,\,w\,\mid\phi_{-n}\lb\lb\phi_{-m},\phi_{n}
\rb,\phi_{m}\rb_{cycl.}\mid\,h\,,\,w\,\rangle\cr
0 &=\ \langle\,h\,,\,w\,\mid\phi_{-n}\lb\lb\phi_{n},\psi_{-m}
\rb_\pm,\psi_{m}\rb_{\pm,cycl.}\mid\,h\,,\,w\,\rangle\cr
0 &=\ \langle\,h\,,\,w\,\mid\psi_{-n}\lb\lb\psi_{-m},\psi_{n}
\rb_\pm,\psi_{m}\rb_{\pm,cycl.}\mid\,h\,,\,w\,\rangle\cr
}\eqno({\rm\rep.5})$$
For a more detailed description of these methods see for example $\q{\wirrep}$.
\mn
\leftline{\bf \res. Results}
\sn
We start with the results for the algebra $\sw(\dh,2)$. This algebra
is most probably the only generically existing
$\sw$-algebra with two generators which
is no Lie superalgebra. The Jacobi identities (\rep.5) lead to
expressions which become trivial if one inserts the relation between the
self coupling constant and $c$. Therefore we assert that this algebra
possesses arbitrary HWRs, an assumption which is supported by the
existence of a free field construction $\q{\komata}$. For the set of
$c$-values $\lbrace -{6\over5},-{39\over4},-{9\over2},{3\over2},33\rbrace$
we constructed two null fields with conformal dimensions
$4$ and $\nh$ in each case (in the case $c=\dh$ with dimensions $4$ and $5$)
$\q{\ich}$. Up to now it is an open question
if a single null field yields enough linearly independent
conditions to restrict the highest weight to a finite number of values.
The value $c = -{6\over5}$ lies in the minimal series of the super Virasoro
algebra and the value $c=-{9\over2}$ belongs to the parabolic cases. Below we
give a general parametrization of the possible $h$-values for these classes.
The remaining $c$-values belong to the class `exceptional', so that we present
the data explicitly ($q = w/C^{\phi}_{\phi\phi}$):
\sn\centerline{\vbox{ \offinterlineskip
\def\tablespace{   height2pt&\omit&&\omit&&\omit&&\omit&\cr }
\def\tablespacesm{ height0.5pt&\omit&&\omit&&\omit&&\omit&\cr }
\def\tablespacefs{ height2pt& \multispan7  &\cr }
\def\tablerule{ \tablespace\noalign{\hrule}\tablespace}
\hrule\halign{&\vrule#&\strut\hskip5pt\hfil#\hfil\hskip5pt\cr
\tablespacefs
& \multispan7 \hfil $\sw(\dh,2)$ \hfil &\cr
\tablespacefs\noalign{\hrule}\tablespace
& $c$ && $-{39\over4}$ && ${3\over2}$ && $33$ &\cr
\tablerule
& \omit && $(h,q)$ &&  $(h,q)$ && $(h,q)$ &\cr
\tablerule
& NS && $\ob(0,0),(-{1\over4},{1\over38})\cb\ \vee$
&& $\ob({1\over2},{1\over2}),({1\over8},{1\over32})\cb\ \vee$
&& $q = {1\over19} h\ \vee$ &\cr
\tablespace
& \omit && $q = {1\over38} (20h+9)$ && $q = -{1\over2} h$
&& $q={2\over19}(2h-3)$ &\cr
\tablerule
& R && $(-{1\over32},{11\over152})\ \vee$
&& $\ob({9\over16},-{9\over32}),({9\over16},{15\over32})\cb\ \vee$
&& $q = {1\over19} h\ \vee$ &\cr
\tablespace
& \omit && $p_0 q^2-p_1 q+ p_2 =0$
&& $q = {1\over32}(3-16h)$
&& $q={2\over19}(2h-3)$ &\cr
\tablespace}\hrule}}
\sn
with $p_0(h) = 92416,\ p_1(h) = 97280 h+31008,\ p_2(h) = 25600h^2+19776h+3861$.
\sn
The first two cases share the same structure of the HWRs. Besides a continuous
one parameter branch of HWRs there further discrete values of $h,q$ exist.
We point out that this pattern appears also for $c$-values of
the other two algebras under investigation with even integer superconformal
dimension of the additional generator. Note that in the exceptional case $c=33$
the restrictions for both sectors coincide.
\sn
Let us now discuss the results of our calculations for
 $\sw(\dh,\delta)$-algebras
which exist for $c$-values in the minimal series of the super
Virasoro algebra. We separate these values into four different series due to
ADE-classification of the modular invariant partition functions built up by
the characters of the correspondent minimal model $\q{\cap}$. For all these
 rational
models one obtains ${\tilde c} = c - 24 h_{min} < \dh$, $h_{min}$ being the
smallest $h$-value of the possible representations.
\sn
The first series consists of the minimal $c$-values with the modular invariant
partition function of the type $(A_{q-1},D_{{p+2}\over2})$. The $\sw$-algebras
belonging to this series have a vanishing self coupling constant.
In complete analogy to the conformal case we can express the characters
of the HWRs of the $\sw(\dh,\delta)$-algebra as a finite sum of superconformal
characters appearing in the modular invariant partition function. In other
words the characters of the $\sw$-algebra diagonalize the partition
function. In the NS sector the characters are given by
(${\cal I}_1 = 2\ \N+1,\ {\cal I}_2 = 2\ \N$):
$$\eqalign{ \chi^{SW}_{i,j} &= \chi_{i,j}+\chi_{i,p-j},
\ \ 1\leq i \leq {\textstyle{q\over2}},
\ 1\leq j \leq {\textstyle{p\over2}}-2,\ i,j\in {\cal I}_1\cr
\chi^{SW}_{i,{p\over2}} &= \chi_{i,{p\over2}}
,\ \ 1\leq i \leq {\textstyle{q\over2}},\ i\in {\cal I}_1\cr}
\eqno({\rm \res.1a})$$
For $i\in {\cal I}_2$ one obtains the characters in the R sector.
Of course one has to
replace $\chi_{i,j}$ by $\hat\chi_{i,j}$.
We stress the fact that the characters span a representation space of the
subgroups of the modular group $SL(2,\Zed)$
generated by the elements $S$ and $T^2$
(NS sector) respectively $ST^2S$ and $T$ (R sector). From (\res.1a) we can
read off immediately the $h$-values of the HWRs in the NS sector
(parametrization due to (\nota.7)):
$$\eqalign{ h^{SW}_{i,j} &= {\rm min}(h_{i,j},h_{i,p-j}),
\ \ 1\leq i \leq {\textstyle{q\over2}},\ 1\leq j \leq
{\textstyle{p\over2}}-2,\ i,j\in {\cal I}_1\cr
h^{SW}_{i,{p\over2}} &= h_{i,{p\over2}}
,\ \ 1\leq i \leq {\textstyle{q\over2}},\ i\in {\cal I}_1\cr}
\eqno({\rm \res.1b})$$
The second series is made up of the minimal $c$-values with a partition
function of the type $(A_{q-1},E_6)$. Realizations are provided by
$\sw(\dh,\sh)$,\ $c={7\over5}$ and $\sw(\dh,\elfh)$,\ $c={10\over7}$ with
non-vanishing self coupling constant. In the NS sector
the characters of the HWRs of the $\sw$-algebra can be written
as follows:
$$\eqalign{ \chi^{SW}_{i,1} &= \chi_{i,1}+\chi_{i,7},
\ \chi^{SW}_{i,2} = \chi_{i,5}+\chi_{i,11},
\ \ 1\leq i \leq {\textstyle{q\over2}},\ i\in {\cal I}_1\cr
\chi^{SW}_{i,3} &= \chi_{i,4}+\chi_{i,8},
\ \ 1\leq i \leq {\textstyle{q\over2}},\ i\in {\cal I}_2\cr }
\eqno({\rm \res.2})$$
The characters in the R sector are obtained by interchanging ${\cal I}_1$
with ${\cal I}_2$.
\sn
The third family contains $c$-values related to the partition functions of
the type $(D_{{q\over2}+1},E_6)$. We found three algebras with vanishing self
coupling constant which fit into this pattern:
$\sw(\dh,\fh)$,\ $c={10\over7}$,\ $\sw(\dh,\nh)$,\
$c={4\over11}$ and $\sw(\dh,\elfh)$,\ $c=-{5\over13}$.
The characters are given by:
$$\eqalign{ \hbox{\rm NS sector:}\ \chi^{SW}_i &= \chi_{i,1}+
\chi_{i,5}+\chi_{i,7}+\chi_{i,11},
\ \ 1\leq i < {\textstyle{q\over2}},\ i\in {\cal I}_1\cr
\chi^{SW}_{q\over2} &= \chi_{{q\over2},1}+\chi_{{q\over2},5}\cr
\hbox{\rm R sector:}\ {\hat\chi}^{SW}_i &=
{\hat\chi}_{i,4}+{\hat\chi}_{i,8},
\ \ 1\leq i < {\textstyle{q\over2}},\ i\in {\cal I}_1\cr
{\hat\chi}^{SW}_{q\over2} &= {\hat\chi}_{{q\over2},4}}
\eqno({\rm \res.3})$$
The last family consist of super minimal $c$-values connected with the
$(A_{q-1},E_8)$ partition functions. Only one $\sw$-algebra with non-vanishing
self coupling constant is constructed so far realizing the series:
\ $\sw(\dh,\elfh)$,\ $c={11\over40}$. The characters read:
$$\eqalign{ \hbox{\rm NS sector:}\ \chi^{SW}_{i,1} &= \chi_{i,1}+
\chi_{i,11}+\chi_{i,19}+\chi_{i,29},
\ \ 1\leq i \leq {\textstyle{q\over2}},\ i\in {\cal I}_1\cr
\chi^{SW}_{i,2} &= \chi_{i,7}+\chi_{i,13}+\chi_{i,17}+\chi_{i,23},
\ \ 1\leq i \leq {\textstyle{q\over2}},\ i\in {\cal I}_1\cr
\hbox{\rm R sector:}\ {\hat\chi}^{SW}_{i,1} &= {\hat\chi}_{i,1}+
{\hat\chi}_{i,11}+{\hat\chi}_{i,19}+{\hat\chi}_{i,29},
\ \ 2\leq i < {\textstyle{q\over2}},\ i\in {\cal I}_2\cr
{\hat\chi}^{SW}_{{q\over2},1} &= {\hat\chi}_{{q\over2},1}+
{\hat\chi}_{{q\over2},11}\cr
{\hat\chi}^{SW}_{i,2} &= {\hat\chi}_{i,7}+{\hat\chi}_{i,13}+
{\hat\chi}_{i,17}+{\hat\chi}_{i,23},
\ \ 2\leq i < {\textstyle{q\over2}},\ i\in {\cal I}_2\cr
{\hat\chi}^{SW}_{{q\over2},2} &=
{\hat\chi}_{{q\over2},7}+{\hat\chi}_{{q\over2},13}}
\eqno({\rm \res.4})$$
\sn
The second class consists of the $\sw$-algebras existing for parabolic
values of $c$. For these rational models we always have ${\tilde c} = \dh$.
We divide this class into two series A and B.
The first series A contains the algebras
with non-vanishing self coupling constant. The central charge is given by
$c=\dh(1-16k)$ with  $4k\in \N$. The superconformal dimension of the additional
generator is $\delta = 8k$. Realizations are provided for
$k={1\over4},{1\over2},{3\over4}$.
The algebras with vanishing self coupling constant belong to the second series
 B.
Here we have $c=\dh(1-16k)$ with $ 2k \in \N $ and  $\delta=3k$.
Realizations are present for $k=1,\dh,2$.
The possible $h$-values can be parametrized with the following formula:
$$h_{r,r} = k(r^2-1),\ \ h_{r,-r} = k(r^2-1) + {\textstyle{\eh}} r^2
\eqno({\rm \res.5})$$
The table below contains the $h$-values of the HWRs for both series. Note
that in the Ramond sector one has to add $1\over16$ to the values given in the
table. For series A in the NS sector the value $h=0$ appears twice,
once with $w=0$ (vacuum) and once with $w\not =0$.
Furthermore we observed that the $h$-values
$h_{0,0},\,h_{1,1},\,h_{1,-1}$ in the R sector
are doubly degenerate.
For series B only one representation is possible with
$w^2\not=0$ in the NS sector,
$h_{\eh,\eh}$ ($k-\lb k\rb=0$) or $h_{\eh,-\eh}$ ($k-\lb k\rb=\eh$).
\sn\centerline{\vbox{ \offinterlineskip
\def\tablespace{ height2pt&\omit&&\omit&&\omit&&\omit &\cr }
\def\tablespacesm{ height0.5pt&\omit&&\omit&&\omit&&\omit &\cr }
\def\tablerule{ \tablespace\noalign{\hrule}\tablespace}
\hrule\halign{&\vrule#&\strut\hskip5pt\hfil#\hfil\hskip5pt\cr
\tablespace
& series && \omit && \hbox{NS sector} && \hbox{R sector} &\cr
\tablerule
& A && $h_{ {l\over{4k}},{l\over{4k}} }$
&& $l = 0,\ldots,4k,8k$ && $l = 0,\ldots,4k$ &\cr
\tablespace
& ($4k\in\N$) && $h_{ {l'\over{4k+2}},-{l'\over{4k+2}} }$
&& $l' = 1,\ldots,4k+1$ && $l' = 0,\ldots,4k+2$ &\cr
\tablerule
& B && $h_{ {l\over{2k}},{l\over{2k}} }$
&& $l = 0,\ldots,\lb k\rb,2k$ && $l = k-\lb k\rb,\ldots,k$ &\cr
\tablespace
& $(2k\in\N)$ && $h_{ {l'\over{2k+1}},-{l'\over{2k+1}} }$
&& $l' = 0,\ldots,\lb k+{\textstyle{\eh}} \rb$
&& $l' = {\textstyle{\eh}}-(k-\lb k\rb),\ldots,k+{\textstyle{\eh}}$ &\cr
\tablespace}\hrule}}
With the knowledge of this data M.\ Flohr was able to give the explicit
form of the characters of the HWRs $\q{\mfl}$.
\sn
The third class is built by the $\sw(\dh,\delta)$-algebras existing for
$c=c(1,s),\ s\in 2\ \N +1$. In complete analogy the conformal case we
obtain infinitely many HWRs lying on one parameter branches. The following
table shows the functional dependence of $w^2$ and $h$:
\sn\centerline{\vbox{ \offinterlineskip
\def\tablespace{ height2pt&\omit&&\omit&&\omit&&\omit &\cr }
\def\tablespacesm{ height0.5pt&\omit&&\omit&&\omit&&\omit &\cr }
\def\tablespacefs{ height2pt&\multispan3 && \multispan3 &\cr }
\def\tablerule{ \tablespace\noalign{\hrule}\tablespace}
\hrule\halign{&\vrule#&\strut\quad\hfil#\hfil\quad\cr
\tablespacefs
& \multispan3 \hfil $\sw(\dh,\fh):\ \ \ c = -{5\over2}$ \hfil
&&  \multispan3 \hfil $\sw(\dh,\nh):\ \ \ c = -{81\over10}$ \hfil  &\cr
\tablespacefs\noalign{\hrule}\tablespace
& \hbox{NS sector} && \hbox{R sector} && \hbox{NS sector} && \hbox{R sector}
 &\cr
\tablespacesm\noalign{\hrule}\tablerule\tablespace
& ${{h^2(6h+1)}\over12}$ && ${{(48h+5)(16h+1)^2}\over24576}$
&& $-{{h^2(5h+2)(10h+3)^2}\over7200}$
&& $-{{(16h+5)^2(80h+9)^2(80h+27)}\over1887436800}$  &\cr
\tablespace\tablespace}\hrule}}
These results support the assumption
of a free field construction for this
class of algebras in analogy to the conformal case $\q{\ka1}$.
\mn
Finally we discuss exceptional $\sw$-algebras. Here the knowledge
of the possible HWRs is of special interest because it is a great help
for the understanding of the structure of these algebras. Especially we are
interested in new rational models with ${\tilde c}$  greater than
$\dh$ and, as we have seen so far, exceptional algebras seem to be the only
candidates.
\sn
At first we consider the cases $\sw(\dh,4)$, $c=-13$ and $\sw(\dh,6)$, $c=-18$,
which have the same mixed structure of their HWRs as the values
$c=-{39\over4}$ and  $c=-\dh$
of $\sw(\dh,2)$. We stress that the conformal dimension of the additional
generator is an even integer in these cases. Here are the concrete results
($ q=w/C^{\phi}_{\phi\phi}$):
\mn\centerline{\vbox{ \offinterlineskip
\def\tablespace{   height2pt&\omit&&\omit&&\omit&&\omit&&\omit&&\omit&\cr }
\def\tablespacesm{ height0.5pt&\omit&&\omit&&\omit&&\omit&&\omit&&\omit&\cr }
\def\tablespacefs{ height2pt&\multispan7 && \multispan3 &\cr}
\def\tablespacehs{ height2pt& \multispan3 && \multispan3 && \multispan3 &\cr }
\def\tablespacehsm{ height0.5pt& \multispan3 && \multispan3 && \multispan3 &\cr
 }
\def\tablerule{ \tablespace \noalign{\hrule} \tablespace }
\hrule\halign{&\vrule#& \strut\quad\hfil#\hfil\quad\cr
\tablespacefs
&  \multispan7 \hfil $\sw(\dh,4):\ \ \ c = -13$ \hfil
&& \multispan3 \hfil $\sw(\dh,6):\ \ \ c=-18$ \hfil  &\cr
\tablespacefs\noalign{\hrule}\tablespacehs
& \multispan3 \hbox{NS sector} && \multispan3 \hbox{R sector}
&&\multispan3 \hbox{NS sector}   &\cr
\tablespacehs\noalign{\hrule}\tablespace
& $h$ && $q$ && $h$ && $q$ && $h$ && $q$ &\cr
\tablespacesm\noalign{\hrule}\tablerule
& $0$ && $0$ && ${1\over8}$ && ${63\over7904}$ && $0$ && $0$ &\cr\tablerule
& $-{1\over6}$ && ${1\over702}$ && $-{1\over24}$ && $-{947\over106704}$
&& $-{1\over4}$ && ${17\over321024}$ &\cr
\tablespace\noalign{\hrule}\tablespacehsm\noalign{\hrule}\tablespacehs
& \multispan3  \hfil $q =  {{538h^2 +549h +137}\over10374}$ \hfil
&& \multispan3 \hfil $p_0(h)\ q^2 + p_1(h)\ q + p_2(h) = 0$ \hfil
&& \multispan3 \hfil $q = {{9076h^3+17404h^2+10827h+2169}\over2588256}$ \hfil
 &\cr
\tablespacehs}\hrule}}
\sn
with
$$\eqalign{
p_0(h) &= 110202753024,\ p_1(h) = -11430322176h^2-8105330688h-1292683392\cr
p_2(h) &= 296390656h^4 + 535662592h^3 + 360220288h^2 + 106998144h + 11899017}$$
Because of the rising complexity of the calculations we have not been able to
 calculate
the restrictions in the R sector for the $\sw(\dh,6)$. However, the results in
 the
NS sector should be sufficient if one has more theoretical knowledge. So far
 there
is no understanding of this mixed behaviour of the HWRs of these algebras.
\sn
For the remaining exceptional cases our calculations revealed the following new
rational models. Because we do not list the $w$-values of the representations
we point out that the $h$-values in the R-sector are degenerate in most cases.
Furthermore we remind the reader of the possibility that some of the $h$-values
may drop out by further investigation because our calculations yield only
necessary conditions.
\mn\centerline{\vbox{ \offinterlineskip
\def\tablespace{ height2pt&\omit&&\omit&&\omit&&\omit&&\omit&\cr }
\def\tablespacesm{ height0.5pt&\omit&&\omit&&\omit&&\omit&&\omit&\cr }
\def\tablespacefs{ height2pt&\multispan9 &\cr }
\def\tablerule{ \tablespace\noalign{\hrule}\tablespace }
\hrule\halign{&\vrule#&\strut\hskip5pt\hfil#\hfil\hskip5pt\cr
\tablespace
& $\delta$ && $c$ && ${\tilde c}$ && sector && $h$  &\cr
\tablespacesm\noalign{\hrule}\tablerule
& $\sh$  && $-{17\over11}$ && ${19\over11}$ && NS &&
 $0$,${1\over22}$,${7\over22}$,${13\over22}$,
${2\over11}$,${25\over22}$,${3\over11}$,$-{1\over11}$,$-{1\over22}$,
$-{3\over22} $&\cr
\tablerule
& \omit  && \omit  && \omit && R &&
 ${3\over88}$,${19\over88}$,${27\over88}$,${51\over88}$,
${59\over88}$,${67\over88}$,${123\over88}$,${131\over88}$,${25\over8}$,
$-{5\over 88}$&\cr
\tablerule
& $4$  && $-{185\over4}$ && ${7\over4}$ && NS && $0$, $-2$, $-{5\over4}$,
 $-{7\over4}$,
$-{11\over6}$, $-{13\over8}$, $-{15\over8}$, $-{19\over12}$ &\cr
\tablerule
& \omit  && \omit && \omit  && R && $-{45\over32}$, $-{51\over32}$,
 $-{61\over32}$,
$-{173\over96}$, $-{185\over96}$, $-{37\over32}$, $-{47\over32}$,
$-{53\over32}$  &\cr
\tablerule
& $4$ && $-{120\over13}$ && ${24\over13}$ && NS && $0$, $-{11\over26}$,
 $-{7\over26}$,
$-{5\over26}$, $-{1\over26}$, $-{6\over13}$, $-{5\over13}$,&\cr
\tablespace
& \omit && \omit&& \omit  && \omit &&  $-{4\over13}$, $-{1\over13}$,
 ${10\over13}$, ${2\over13}$,
${55\over26}$, ${23\over26}$, ${9\over26}$, ${3\over26}$ &\cr
\tablerule
& \omit && \omit&& \omit  && R && $4$, $20\over13$, $18\over13$, $17\over13$,
 $12\over13$,
$6\over13$, $3\over13$, $2\over13$, &\cr
\tablespace
& \omit && \omit&& \omit  && \omit && $1\over13$, $-{2\over13}$, $-{3\over13}$,
$-{4\over13}$, $-{5\over13}$  &\cr
\tablerule
& $\elfh$ && $-{705\over8}$ && ${15\over8}$ && NS && $0$, $-{7\over2}$,
 $-{11\over3}$,
$-{13\over4}$, $-{15\over4}$, $-{15\over8}$, &\cr
\tablespace
& \omit && \omit && \omit && \omit && $-{23\over8}$, $-{25\over8}$,
$-{29\over8}$, $-{29\over12}$, $-{79\over24}$, $-{85\over24}$ &\cr
\tablerule
& \omit && \omit && \omit && R && $-{159\over64}$, $-{191\over64}$,
 $-{199\over64}$,
$-{211\over64}$, $-{219\over64}$, $-{227\over64}$, &\cr
\tablespace
& \omit && \omit&& \omit  && \omit && $-{231\over64}$,
$-{235\over64}$, $-{593\over192}$, $-{605\over192}$, $-{641\over192}$,
$-{701\over192}$ &\cr
\tablerule
& $\elfh$ && $-{155\over19}$ && ${37\over19}$ && NS && $0$, $-{3\over19}$,
 $-{5\over19}$,
$-{7\over19}$, $-{8\over19}$, $-{3\over38}$, $-{5\over38}$, $-{11\over38}$,
&\cr
\tablespace
& \omit && \omit&& \omit  && \omit && $-{13\over38}$, $-{15\over38}$,
${1\over19}$, ${6\over19}$, ${10\over19}$,
${17\over19}$, ${26\over19}$, ${70\over19}$, ${11\over38}$, ${17\over38}$,
${29\over38}$, ${75\over38}$ &\cr
\tablerule
& $6$ && $-{2241\over20}$ && ${39\over20}$ && NS &&
$0$, $-{9\over2}$, $-{14\over3}$, $-{17\over4}$,
$-{19\over4}$, $-{21\over5}$, $-{22\over5}$,&\cr
\tablespace
&\omit && \omit && \omit && \omit &&
$-{39\over10}$, $-{47\over12}$, $-{53\over20}$, $-{89\over20}$,
 $-{93\over20}$, $-{137\over30}$, $-{229\over60}$ &\cr
\tablespace}\hrule}}
\sn
In the case $\sw(\dh,\elfh)$, $c=-{155\over19}$ our
calculations in the R sector yielded so far
only one relation between $h$ and $q$. Due to computational problems
we have not been able to evaluate more complicated Jacobi identities
and therefore the $h$-values of the model cannot be presented.
As mentioned above the computational
problems in the R sector were enormous for the $\sw(\dh,6)$ and forced us
to stop the calculations here.
\sn
\leftline{\bf \summa. Summary}
\sn
In this paper we investigated the highest weight representations of
the $N=1$ $\sw$-algebras with two generators for dimensions of the
additional superprimary between two and six. For the algebras existing
for discrete values of the central charge $c$ we studied Jacobi
identities. Our calculations yield necessary
conditions for consistent highest weight representations. We stress
the fact that in the minimal and parabolic cases
the HWRs obtained this way coincide exactly with the expected values.
This is a strong hint that in the other cases the given HWRs do indeed
exist. For the generically existing algebra $\sw(\dh,2)$ the Jacobi
identities are trivial so that arbitrary HWRs are possible.
Imposing the physical condition that
null fields vanish in all HWRs we found for special $c$-values
models fitting in the general pattern.
\sn
For the algebras existing for $c$-values contained in the minimal series
of the super Virasoro algebra the allowed $h$-values are exactly those
which are obtained if one assumes that the characters of the possible HWRs
of the $\sw$-algebra are finite sums of superconformal characters and
diagonalize the modular invariant partition function belonging to this
$c$-value according to the ADE-classification. Furthermore we observed that
in the cases where two different modular invariant partition functions
are possible for a single $c$-value there indeed exist two different
$\sw(\dh,\delta)$-algebras diagonalizing both solutions
of A.\ Cappelli et al. Looking at the partition functions one may
conjecture that the algebra with the higher conformal dimension of
the simple additional superprimary is a subalgebra of the other
algebra existing for the special $c$-value. This fact has already been
pointed out in $\q{\hornfeck}\q{\supwir}$. Our results
suggest that for every modular invariant partition function of
A.\ Cappelli et al. there exists a $\sw(\dh,\delta)$-algebra
diagonalizing it.
\sn
As expected we obtained for the parabolic algebras rational models
with $\tilde c =\dh$. We also gave a parametrization of
the $h$-values. After the presence of this data the characters
of the HWRs were given in $\q{\mfl}$.
For the $\sw(\dh,\delta)$-algebras existing for $c=c(1,s)$ with
$\delta = s-\eh$, $2\leq s\in\N$ our results propose infinitly
many HWRs lying on one parameter branches in analogy to the
conformal case $\q{\ka1}\q{\wirrep}$.
\sn
For the $\sw(\dh,\delta)$-algebras with
$\delta\in\lbrace 2,4,6\rbrace$ we found a new interesting
structure of their HWRs for special $c$-values. Here one
has a mixed behaviour: besides a continuous one parameter
branch HWRs with discrete values of the highest
weight exist.

\noindent
The remaining exceptional algebras lead to new rational
models with $2 > \tilde c > \dh$. With this data the
first step to a complete understanding of these cases is
done.
\sn
\leftline{\bf Acknowledgements}
\sn
We are very grateful to W.\ Nahm, R.\ Blumenhagen, M.\ Flohr,
J.\ Kellendonk, S.\ Mallwitz, A.\ Recknagel, M.\ R{\"o}sgen,
M.\ Terhoeven and R.\ Varnhagen for many useful discussions.
It is a pleasure to thank the Max-Planck-Institut f{\"u}r
Mathematik in Bonn-Beuel especially M.\ Auferkorte, Th.\ Berger
and S.\ Maurmann because the greatest part of the calculations
has been performed there.
\bn
\leftline{{\bf References}}
\mn
\settabs\+&\phantom{---------}
&\phantom
{------------------------------------------------------------------------------}
& \cr
\+ &$\q{\bbss}$& F.A.\ Bais, P.\ Bouwknegt, M.\ Surridge, K.\ Schoutens & \cr
\+ &           & {\it Extensions of the Virasoro Algebra Constructed form
 Kac-Moody Algebras Using} & \cr
\+ &           & {\it Higher Order Casimir Invariants}, Nucl.\ Phys.\ {\bf
B304}
 (1988) p.\ 348 & \cr
\+ &$\q{\bai}$ & F.A.\ Bais, P.\ Bouwknegt, M.\ Surridge, K.\ Schoutens & \cr
\+ &           & {\it Coset Construction for Extended Virasoro Algebras} & \cr
\+ &           & Nucl.\ Phys.\ {\bf B304} (1988) p.\ 371 & \cr
\+ &$\q{\bal}$ & J.\ Balog, L.\ Feh\'er, P.\ Forg\'acs, L.\ O'Raifeartaigh, A.\
 Wipf & \cr
\+ &           & {\it Kac-Moody Realization of $\w$-Algebras} & \cr
\+ &           & Phys.\ Lett.\ {\bf B244} (1990) p.\ 435 & \cr
\+ &$\q{\bpz}$ & A.A.\ Belavin, A.M.\ Polyakov, A.B.\ Zamolodchikov & \cr
\+ &           & {\it Infinite Conformal Symmetry in Two-Dimensional Quantum
 Field Theory}  & \cr
\+ &           & Nucl.\ Phys.\ {\bf B241} (1984) p.\ 333  & \cr
\+ &$\q{\bershadsky}$
               & M.A. Bershadsky, V.G. Knizhnik, M.G. Teitelman& \cr
\+ &           & {\it Superconformal Symmetry in Two Dimensions } & \cr
\+ &           & Phys.\ Lett.\ {\bf B151} (1985) p.\ 31 & \cr
\+ &$\q{\blg}$ & A.\ Bilal, J.L.\ Gervais & \cr
\+ &           & {\it Systematic Construction of Conformal Theories with
 Higher-Spin Virasoro} & \cr
\+ &           & {\it Symmetries}, Nucl.\ Phys.\ {\bf B318} (1989) p.\ 579 &
\cr
\+ & $\q{\blumenhagen}$
               & R.\ Blumenhagen, {\it Covariant Construction of $N=1$ Super
 $\w$-Algebras } & \cr
\+ &           & Preprint BONN-HE-91-20 (1991),\ \ to appear in Nucl.\ Phys.\
 {\bf B} & \cr
\+ &$\q{\supwir}$
               & R.\ Blumenhagen, W.\ Eholzer, A.\ Honecker, R.\ H{\"u}bel &
\cr
\+ &           & {\it New N=1 Extended Superconformal Algebras with Two and
 Three Generators } & \cr
\+ &           & Preprint BONN-HE-92-02 (1992),\ \ to appear in Int.\ Jour.\
 Mod.\ Phys.\ {\bf A} & \cr
\+ &$\q{\blm}$ & R.\ Blumenhagen, M.\ Flohr, A.\ Kliem, W.\ Nahm, A.\
Recknagel,
 R.\ Varnhagen & \cr
\+ &           & {\it $\w$-Algebras with Two and Three Generators}, Nucl.\
 Phys.\ {\bf B361} (1991) p.\ 255 & \cr
\+ &$\q{\bou}$ & P.\ Bouwknegt, {\it Extended Conformal Algebras}, Phys.\
Lett.\
 {\bf B207} (1988) p.\ 295 & \cr
\+ &$\q{\cap}$ & A.\ Cappelli, C.\ Itzykson, J.B.\ Zuber & \cr
\+ &           & {\it The A-D-E Classification of Minimal and $A_1^{(1)}$
 Conformal Invariant Theories} & \cr
\+ &           & Comm.\ Math.\ Phys.\ 113 (1987) p.\ 1 & \cr
\+ &$\q{\wirrep}$
               & W.\ Eholzer, M.\ Flohr, A.\ Honecker, R.\ H{\"u}bel, W.\ Nahm,
 R.\ Varnhagen  & \cr
\+ &           & {\it Representations of $\w$-Algebras with Two Generators and
 New Rational Models } & \cr
\+ &           & Preprint BONN-HE-91-22 (1991),\ \ to appear in Nucl.\ Phys.\
 {\bf B} & \cr
\+ & $\q{\eichenherr}$
               & H.\ Eichenherr, {\it Minimal Operator Algebras in
 Superconformal Quantum Field Theory } & \cr
\+ &           & Phys.\ Lett.\ {\bf B151} (1985) p.\ 26 & \cr
\+ &$\q{\fal}$ & V.A.\ Fateev, S.L.\ Lyk'anov & \cr
\+ &           & {\it The Models of Two Dimensional Quantum Conformal Field
 Theory with $\Zed_n$ Symmetry} &\cr
\+ &           & Int.\ Journ.\ Mod.\ Phys.\ {\bf A3} (1988) p.\ 507 &\cr
\+ & $\q{\fateev}$
               & V.A.\ Fateev, A.B.\ Zamolodchikov & \cr
\+ &           & {\it Conformal Quantum Field Theory Models in Two Dimensions
 Having} & \cr
\+ &           & {\it $\Zed_3$ Symmetry}, Nucl.\ Phys.\ {\bf B280} (1987) p.\
 644 & \cr
\+ &$\q{\fig}$ & J.M.\ Figueroa-O'Farrill, S.\ Schrans, {\it The Spin 6
Extended
 Conformal Algebra} & \cr
\+ &           & Phys.\ Lett.\ {\bf B245} (1990) p.\ 471 & \cr
\+ & $\q{\jfss}$
               & J.M.\ Figueroa-O'Farrill, S.\ Schrans,  {\it The Conformal
 Bootstrap and Super $\w$ Algebras } & \cr
\+ &           & Int.\ Jour.\ Mod.\ Phys.\ {\bf A7} (1992) p.\ 591 & \cr
\+ &$\q{\mfl}$ & M.\ Flohr, {\it $\w$-Algebras, New Rational Models and
 Completeness of the $c=1$ Classification} &\cr
\+ &           & Preprint BONN-HE-92-08 (1992) & \cr
\+ & $\q{\goddard}$
               & P.\ Goddard, A.\ Kent, D.\ Olive & \cr
\+ &           & {\it Unitary Representations of the Virasoro and
Super-Virasoro
 Algebra}  & \cr
\+ &           & Comm.\ Math.\ Phys.\ 103 (1986) p.\ 105 & \cr
\+ &$\q{\gko}$ &  P.\ Goddard, A.\ Kent, D.\ Olive &\cr
\+ &           &  {\it Virasoro Algebras and Coset Space Models} & \cr
\+ &           &   Phys.\ Lett.\ {\bf B152} (1985) p.\ 88 & \cr
\+ & $\q{\hornfeck}$  & K.\ Hornfeck, {\it Realizations for the ${\cal
 SW}({7\over2})$-Algebra} & \cr
\+ &           & {\it and the Minimal Supersymmetric Extension of ${\cal WA}_3$
 } & \cr
\+ &           & Preprint London, King's College 91.08.26 & \cr
\+ & $\q{\ich}$ & R.\ H{\"u}bel, {\it Darstellungstheorie von $\w$- und
 Super-$\w$-Algebren} &\cr
\+ &            & Diplomarbeit BONN-IR-92-11 (1992) &\cr
\+ & $\q{\inami}$ & T.\ Inami, Y.\ Matsuo, I.\ Yamanaka & \cr
\+ &           & {\it Extended Conformal Algebras with $N=1$ Supersymmetry} &
 \cr
\+ &           & Phys.\ Lett.\ {\bf B215} (1988) p.\ 701 & \cr
\+ &$\q{\ka1}$ & H.G.\ Kausch & \cr
\+ &           & {\it Extended Conformal Algebras Generated by a Multiplet of
 Primary Fields} & \cr
\+ &           & Phys.\ Lett.\ {\bf B259} (1991) p.\ 448 & \cr
\+ &$\q{\kau}$ & H.G.\ Kausch, G.M.T.\ Watts, {\it A Study of $\w$-Algebras
 Using Jacobi Identities}& \cr
\+ &           & Nucl.\ Phys.\ {\bf B354} (1991) p.\ 740 & \cr
\+ &$\q{\komata}$ & S.\ Komata, K.\ Mohri, H.\ Nohara &\cr
\+ &           & {\it Classical and Quantum Extended Superconformal Algebra}
 &\cr
\+ &           & Nucl.\ Phys.\ {\bf B359} (1991) p.\ 168 &\cr
\+ & $\q{\mussardoi}$
               & G.\ Mussardo, G.\ Sotkov, M.\ Stanishkov & \cr
\+ &           & {\it Ramond Sector of Supersymmetric Minimal Models }& \cr
\+ &           & Phys.\ Lett.\ {\bf B195} (1987) p.\ 397 & \cr
\+ & $\q{\mussardoii}$
               & G.\ Mussardo, G.\ Sotkov, M.\ Stanishkov & \cr
\+ &           & {\it Fine Structure of Supersymmetric Operator Product
 Expansion Algebras} & \cr
\+ &           & Nucl.\ Phys.\ {\bf B305} (1988) p.\ 69 & \cr
\+ &$\q{\nahm}$
               & W.\ Nahm, {\it Chiral Algebras of Two-Dimensional Chiral Field
 Theories and Their} & \cr
\+ &           & {\it Normal Ordered Products  }, Proceedings Trieste
Conference
 on & \cr
\+ &           & Recent Developments in Conformal Field Theories, ICTP, Trieste
 (1989) p.\ 81 & \cr
\+ &$\q{\scheunert}$
               & M.\ Scheunert, {\it The Theory of Lie Superalgebras} &\cr
\+ &           & Lecture Notes in Mathematics 716 (1979) &\cr
\+ &$\q{\rva}$ & R.\ Varnhagen, {\it Characters and Representations of New
 Fermionic $\w$-Algebras} & \cr
\+ &           & Phys.\ Lett.\ {\bf B275} (1992) p.\ 87 & \cr
\+ &$\q{\zam}$ & A.B.\ Zamolodchikov & \cr
\+ &           & {\it Infinite Additional Symmetries in Two-Dimensional
 Conformal Quantum Field}  & \cr
\+ &           & {\it Theory}, Theor.\ Math.\ Phys.\ 65 (1986) p.\ 1205  & \cr
\vfill
\end